\documentclass[]{raa}            
\usepackage{graphicx,times}
\usepackage{natbib}

\providecommand{\bm}[1]{\mbox{\boldmath$#1$\unboldmath}}

\begin{document}

\title{Spectroscopic and Photometric Observations of Unidentified Ultraviolate Variable 
Objects in GUVV-2 Catalog}

 \volnopage{ {\bf 0000} Vol.\ {\bf 0} No. {\bf XX}, 000--000}
   \setcounter{page}{1}

   \author{Li, Y
      \inst{1,2}
   \and Wang, J
      \inst{2}
   \and Wei, J. Y. 
      \inst{2}
   \and He, X. T.
      \inst{1}
   }
   \institute{Department of Astronomy, Beijing Normal University, Beijing, 100875, China; 
              {\it ollie-321@hotmail.com}\\
        \and National Astronomical Observatories, Chinese Academy of Sciences,
             Beijing 100012, China\\
\vs \no
   {\small Received [year] [month] [day]; accepted [year] [month] [day] }
}

\abstract{An NUV-optical diagram made for sources from the secend \it Galaxy Evolution
Explorer \rm (\it GALEX\rm\,) Ultraviolet Variability (GUVV-2) Catalog provide us
a method to tentatively classify the unknown GUVV2 sources by their NUV-optical magnitudes.
On the purpose of testing the correctness and generality of the method, we carry out a program
on the spectroscopic observations of the unidentified GUVV2 sources.
The spectroscopic identification for these
37 sources are 19 type -A to -F stars, 10 type -G to -K stars and 7 M dwarf
stars together with an AGN. We also present the light curves in R-band for two RR
Lyrae star candidates selected from the NUV-optical diagram, both of which perform cyclic
variations. Combining there light curves and colors, we classify them as RR Lyrae stars.
To confirm the results, we shows a color-color diagram for the 37 newly spectroscopically identified
objects compared with the previously identified ones, which manifests good consistence
with our former results, indicating that the ultroviolet variable sources can be 
initially classified by their NUV/optical color-color diagram.
\keywords{stars: variables: general --- galaxies: active --- methods: observational --- ultraviolet emission
}
}

   \authorrunning{Li et al.}           
   \titlerunning{Spectroscopic and Photometric Observations of Unidentified GUVV2 Sources} 
   \maketitle

\section{Introduction}           
\label{sect:intro}

The NASA \it Galaxy Evolution Explorer \rm (\it GALEX\rm\,) satellite (Martin et al. 2005) is an ongoing 
mission with the goal to study star formation in galaxies and its evolution with time in ultraviolet 
bands. A modified Ritchey-Chretien telescope with diameter of 50 cm and field-of-view of 1.2\symbol{23}
is used to image and spectroscopically observe sky in two ultraviolet bands (NUV 1750--2750 \AA, FUV 1350--1750 \AA)
down to an AB magnitude of  $m_{\mathrm{AB}}\sim25$ in the deepest modes (Morrissey et al. 2007).
It has totally imaged 2/3 sky\footnote{The \it GALEX\rm\, pointings are kept away from the 
Galactic plane and Magellanic clouds in order to avoid the damage of its detector caused by the bright stars 
($\sim10$th magnitude)} during its All-sky Imaging Survey (AIS) and during its Medium Imaging Survey (MIS) and Deep 
Imaging Survey (DIS) since the launch on 2003 April 28. 

Repeated observations of selected areas of the sky enabling numerous sources to have their FUV and NUV fluxes
to be determined at many different epochs. Thus, the variable ultraviolet sources can be detected and many of
which exhibit much larger amplitudes of variation in the ultraviolet region than that typically found at
visible wavelengths (Wheatley et al. 2008).

In addtion to the shallow survey, \it GALEX \rm\, repeatly perfomred the deeper image observations for
the selected sky area. The repeat observations in DIS and MIS allow Welsh et al. (2005) released the
first \it GALEX \rm\, ultraviolet variability catalog (GUVV1), which contains 84 variable and transient UV sources
obtained during the first 15 months of the \it GALEX\rm\, all-sky imaging survey. M dwarf flare stars and ab-type RR Lyrae stars
are the largest ultraviolet variable objects detected by \it GALEX\rm\, in GUVV1 (Welsh et al. 2005).
The sample is subsequently enlarged to the second version (GUVV2, Wheatley et al. 2008) by the repeated
observations in DIS and in certain Guest Investigator (GI) observations. GUVV2
contains 410 variable souces obtained during the period 2003 June - 2006 June. About 72\% sources
in GUVV2 hade not been spectroscopically identified at that time. The already identifid objects by SIMBAD catalog
in GUVV2 are mainly quasars/AGNs (77), RR Lyrae stars (6), dMe stars(2) and X-ray binaries (19). 

We systematically performed follow-up optical spectroscopic and photometric observations for the
unknown sources listed in the GUVV2 catalog. 

In this paper, we provide a method to tentatively classify ultroviolet variable sources into
several common source types by their NUV-optical color-color diagram.
This method is confirmed by our spectroscopic and photometric observations. 
We report results of the spectroscopic identification for 37 previously unkonwn sources and 
photometric monitors carried out for two unidentified RR Lyrae star candidates to confirm their variability
through their light curves. 
The paper is orgnized as follows. \S 2 presents the motivations, observations and data reductions. The 
results and discussions are given in \S 3 and \S 4, respectively, and followed by the conclusion.

\section{Motivations, Observations and Data Reduction}
\label{sect:Obs}
\subsection{Motivations}
In GUVV2, of the 114 sources with previous SIMBAD catalog identifications, there are three mainly common types:
quasars/AGNs, RR Lyrae stars, M dwarf stars. In order to statistically analyze the spectral types of GUVV2 sources,
we have produced a color-color diagram for them by using their NUV magnitudes and optical magnitudes of B2 and R2
from USNO-B1.0 catalog (Monet et al. 2003) (Fig. 1). 
The sources contained in this diagram are previously known quasars/AGNs,
RR Lyrae stars, M dwarf stars and the unidentified ones.
The plot of ($NUV_{max}$-B2) versus (B2-R2) magnitudes reveals that the sources are segregated into different isolated
regions according to their spectral types. The AGNs are mainly clustered at the bottom-left corner because of their
non-thermal blue spectrum and strong NUV emission from the hot disk arround the central supermassive black hole.
The RR Lyare stars are located within the region at right of the AGNs bacause of the strong Balmer jumps.
The M dwarf stars are located at the top-right end of the diagram simply because of their low stellar surface temperature.
The distribution of these previously known sources allows us to classify the unknown sources simply by their
NUV-optical colors in advance.

To preliminarily test the results, we have chosen two RR Lyrae star candidates from the region where RR Lyrae
stars locate to perform photometric monitors. These two sources are also selected as RR Lyrae star candidates
by their SDSS magnitudes (Wheatley et al. 2008).

Taking account of the poor sample of the GUVV2-SIMBAD crossed sources and their potential bias, we have performed further
spectroscopic observations for the randomly chosen unknown sources in GUVV2 to verify the correctness and generality
of this method. 

\begin{figure}[h!!!]
   \centering
   \includegraphics[width=9.0cm, angle=0]{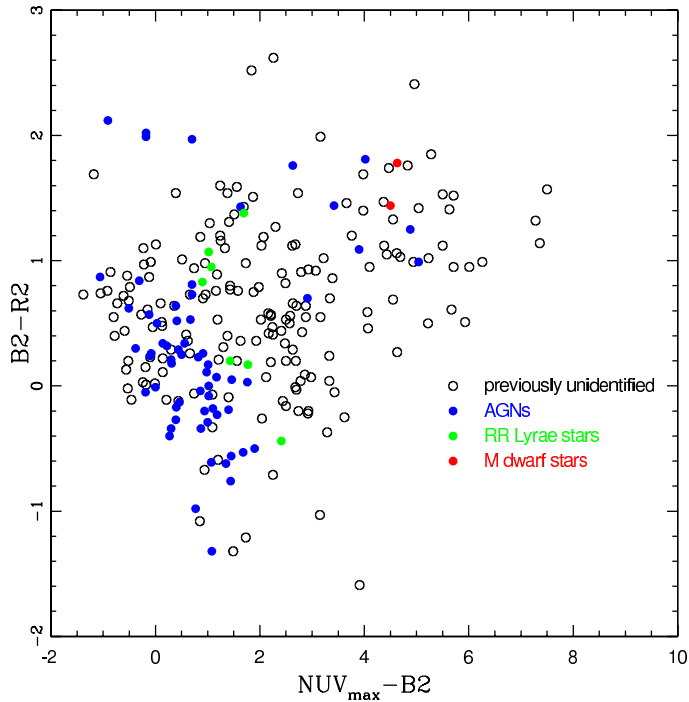}

   \begin{minipage}[]{85mm}
  \caption{Plot of ($NUV_{max}- B2$) vs. (B2 -- R2) for previously identified variable sorces.
 The filled circles with colors of blue, green and red are previosly identified AGNs, RR Lyrae stars
and M dwarf stars, respectively. The black open circles have no current identifications. }

   \end{minipage}
   \label{Fig1}
   \end{figure}

\subsection{Spectroscopic observations}

We only carryed out the spectroscopic observations for the unidentified GUVV2 sources with B-band
magnitude brighter than 17.5 magnitude and declination $>$ $-15^\circ$ due to the constraint of the 
observatory site and the instrumental capability. Basing on the limitation of our observing time, sources were
chosen according to their suitable right ascention in every run of our observations. This principle results in
a list of 37 unidentified sources to be observed (Table 1.). 

The spectroscopic observations were performed in several observing runs between 2009 and 2010
by using the National Astronomical Observatories, Chinese Academy of Sciences (NAOC) 2.16m telescope 
in Xinglong Observatory. The observations were taken with the OMR spectrograph
attached at the Cassegrain focus, and a back-illuminated SPEC10 $1024\times400$ CCD camera employed as the detector. 
A grating of 300 g mm$^{-1}$ and a slit of around $2\arcsec.0$  oriented in the south-north direction 
were used. This setup results in a final spectral resolution of $\sim$9\AA\ as 
measured from both comparison spectra and night sky emission lines. 
The spectroscopic observation log for the identified objects was listed in Table 1. 

The wavelength calibrations were carried out by the He/Ne/Ar comparison
arcs taken several times in each observation run. Two or three Kitt Peak National Observatory (KPNO)
standard stars (Massey et al. 1988) were observed per night for flux calibration.

The raw data were bias subtracted, flat-field corrected, and cosmic-rays removed by using the IRAF 
package\footnote{IRAF is distributed by the National Optical Astronomy Observatory, which
is operated by the Association of Universities for Research in Astronomy, Inc., under cooperative 
agreement with the National Science Foundation; http://iraf.noao.edu.}.
The one-dimensional sky-subtracted spectra were then wavelength and flux calibrated. 
The uncertainties of the wavelength and flux calibrations are no more than 1\AA\ and 20\%, respectively.

\subsection{Photometric observations}

To confirm the results, on the basis of our observing time, we have chosen two unidentified RR Lyrae star candidates
from the region where RR Lyrae stars locates on the NUV-optical diagram with suitable brightness and right ascention to
perform photometric monitors. These two sources (GALEX\,J100133.2+014328.5 and GALEX\,J155800.6+535233.6) are
also selected as RR Lyrae star candidates by their SDSS magnitudes (Wheatley et al. 2008).
Each of the two sources was monitored by using the 85cm NBT telescope at Xinglong Observatory of NAOC
for two days from 2010 March 09 to 2010 March 10.
A standard Johnson-Cousin-Bessel system is mounted on the primary focus of the telescope.
A PI1024 BFT $1024\times1024$ CCD is used as the detector. The field-of-view is about 
$16.5\times16.5\ \mathrm {arcmin^2}$ at a focal ratio of 3.27 ($f$=2780 mm), which results in 
an image scale of 0.96 arcsec per pixel (Zhou et al. 2009). The standard Johnson V- and R-band filters 
were used during our observations. The typical exposure times for the V- and R-band filters is 800s and 600s, 
respectively. In all the observations, the sky flat-field frames in U, V and R passbands were
obtained before and after each observation run during the twilight time.

The raw CCD images were preliminarily reduced through the standard CCDPROC routines in the IRAF package.
For the purpose of differential photometry, we have first chosen several bright stars with good seeing
from the same CCD frame as comparison stars. Then some check stars with brightness comparable to the object
are selected to assess the errors in photometry.
Aperture photometry is adopted to calculate the instrumental magnitudes of the objects 
and the selected stars by the APPHOT task in the IRAF package.

\begin{table}
\bc

\begin{minipage}[]{200mm}

\caption[]{ The Observation Log For The identified Objects}\end{minipage}

\tiny

 \begin{tabular}{cccccccccccc}
  \hline\noalign{\smallskip}
           & $\alpha$ & $\delta$ &         &           &           &      &      &     &           & Exposure &              \\
  GALEX ID &(J2000.0)&(J2000.0)  &$N_{det}$&$NUV_{max}$&$\Delta NUV$& $B2$& $R2$ & $I$ & Obs.Date  & (sec)    & Spectral Type\\
  (1)      & (2)      & (3)      &    (4)  &    (5)    & (6)       & (7)  & (8)  & (9) &    (10)   &  (11)    & (12)         \\
  \hline\noalign{\smallskip}
GALEX J002442.4+165808.0 &00 24 42.52   &+16 58 06.4& 8& 17.68 &0.71& 13.62 & 13.03 & 12.67 & Oct. 14 2009         & 900   & F  \\
GALEX J004215.3+200957.3 &00 42 15.36   &+20 09 59.8&15& 18.94 &1.65& 13.88 & 12.35 & 11.08 & Oct. 14 2009         & 900   & G  \\
GALEX J013642.0-062743.1 &01 36 42.03   &-06 27 43.8&10& 19.70 &0.63& 16.79 & 16.31 & 16.12 & Oct. 14 2009  & 2400  & A \\
GALEX J021547.1-045207.8 &02 15 47.14   &-04 52 08.2& 7& 19.60 &0.67& 15.06 & 12.96 & 12.37 & Sep. 21 2009  & 900   & G \\
GALEX J022003.0-032910.7 &02 20 03.08   &-03 29 10.1& 7& 20.01 &0.73& 16.19 & 14.44 & 13.77 & Oct. 14 2009  & 1800  &  K \\
GALEX J022050.6-061528.0 &02 20 50.71   &-06 15 28.4& 7& 18.99 &0.77& 14.67 & 13.14 & 12.24 & Sep. 21 2009  & 900   &  G \\
GALEX J042733.6+165222.2 &04 27 33.67   &+16 52 21.8& 5& 20.18 &0.73& 16.34 & 14.90 & 12.93 & Jan. 23 2010  & 900   &  M \\ 
GALEX J042832.3+165821.6 &04 28 32.37   &+16 58 21.2& 7& 19.30 &1.64& 16.60 & 15.96 & 14.90 & Jan. 23 2010         & 1800  & F  \\
GALEX J043431.2+172220.1 &04 34 31.28   &+17 22 20.2&13& 19.66 &0.71& 17.22 & 15.05 & 14.36 & Jan. 25 2010  & 2400  &  M \\
GALEX J081226.4+033320.5 &08 12 26.37   &+03 33 20.4&77& 18.95 &1.95& 17.26 & 15.88 & 16.00 & Jan. 25 2010         & 3600  &  F \\ 
GALEX J090853.4-021242.8 &09 08 53.48   &-02 12 43.2&13& 19.23 &0.98& 15.90 & 15.66 & 16.16 & Nov. 10 2010         & 3600  &  F \\  
GALEX J091332.6-101108.2 &09 13 32.65   &-10 11 08.1&15& 19.11 &1.47& 15.35 & 14.15 & 13.46 & Nov. 10 2010         & 2400  &  AGN\\
GALEX J095847.6+021815.3 &09 58 47.64   &+02 18 15.1&11& 18.62 &1.42& 14.59 & 13.54 & 13.03 & Mar. 12 2010  & 1200  &  F \\ 
GALEX J102841.9+570840.4 &10 28 41.94   &+57 08 40.1&24& 18.27 &1.26& 14.01 & 12.85 & 12.49 & Mar. 15 2010  & 900   &  G \\ 
GALEX J102911.8+575806.1 &10 29 11.81   &+57 58 05.5&18& 19.54 &1.13& 11.52 &  9.78 &  8.96 & Mar. 15 2010  & 600   &  M \\ 
GALEX J103538.2+581549.1 &10 35 38.19   &+58 15 49.2&20& 15.53 &1.92& 13.59 & 12.96 & 12.24 & Mar. 15 2010  & 700   &  A \\ 
GALEX J120519.8-075605.6 &12 05 19.83   &-07 56 06.1&31& 19.63 &1.19& 15.20 & 13.41 & 12.58 & Mar. 13 2010  & 1200  &  K \\
GALEX J123523.7+614132.4 &12 35 23.71   &+61 41 31.9&92& 20.34 &0.61& 14.50 & 13.48 & 12.84 & Mar. 13 2010  & 900   &  K \\ 
GALEX J123718.3+624207.9 &12 37 18.39	&+62 42 07.9&29& 20.23 &0.75& 13.89 & 12.18 & 10.81 & Jan. 23 2010	 & 2400  &  M \\ 
GALEX J132602.6+273502.3 &13 26 02.68   &+27 35 02.1&70& 17.40 &3.33& 14.66 & 12.10 & 10.55 & Jan. 23 2010  & 1800  &  M \\
GALEX J143126.4+342710.3 &14 31 26.36   &+34 27 10.4&18& 18.40 &0.86& 13.78 & 12.72 & 11.68 & Mar. 15 2010  & 600   &  G \\ 
GALEX J145110.3+310639.8 &14 51 10.45   &+31 06 40.7&24& 18.72 &0.86& 13.58 & 11.24 &  9.78 & Mar. 15 2010  & 600   &  M \\ 
GALEX J160902.8+524224.4 &16 09 02.81   &+52 42 24.4&18& 19.11 &0.74& 17.36 & 18.00 & 17.44 & Oct. 05 2010  & 3600  &   A\\
GALEX J163853.2+413932.7 &16 38 53.25   &+41 39 36.1&29& 19.97 &1.02& 12.33 & 11.34 & 10.93 & Oct. 06 2010         & 300   & K \\ 
GALEX J163952.7+420951.6 &16 39 52.77   &+42 09 52.0&19& 18.29 &1.75& 15.79 & 15.95 & 15.64 & Oct. 05 2010         & 1200  & A  \\ 
GALEX J171250.9+582748.6 &17 12 50.87   &+58 27 48.4&14& 18.51 &0.63& 16.06 & 15.74 & 15.40 & Oct. 06 2010  & 900   & A  \\ 
GALEX J172033.4+585513.2 &17 20 33.51   &+58 55 13.5&37& 20.12 &0.80& 15.12 & 14.34 & 14.13 & Oct. 06 2010  & 900   &  G \\ 
GALEX J194443.4-074959.5 &19 44 43.40   &-07 49 59.4&11& 19.85 &0.76& 15.08 & 12.90 & 11.36 & Sep. 20 2009         & 600   &   M\\
GALEX J194704.9-075446.1 &19 47 04.99   &-07 54 47.0& 7& 19.56 &1.43& 16.78 & 16.74 & 16.32 & Jun. 30 2009  & 1800  &  F \\
GALEX J203922.4-010346.0 &20 39 22.45   &-01 03 46.1&19& 16.88 &0.81& 14.54 & 14.04 & 13.39 & Sep. 21 2009  & 600   &  A \\ 
GALEX J203958.8-010714.3 &20 39 58.85   &-01 07 14.6&19& 18.43 &1.20& 16.33 & 14.93 & 14.36 & Oct. 06 2010  & 1200  &  F \\ 
GALEX J204024.3-010950.0 &20 40 24.30   &-01 09 50.7&15& 19.06 &0.72& 16.53 & 16.29 & 15.35 & Jun. 30 2009         & 1800  &  F \\ 
GALEX J204114.7-005220.9 &20 41 14.80   &-00 52 21.0&19& 19.59 &1.38& 17.69 & 16.50 & 15.96 & Jun. 30 2009         & 1800  &  F \\ 
GALEX J214909.5-050558.6 &21 49 09.51   &-05 05 59.1&26& 18.67 &2.28& 17.29 & 16.89 & 16.50 & Oct. 13 2009  & 2400  &  F \\ 
GALEX J223601.0+134711.5 &22 36 01.10   &+13 47 11.2&27& 19.16 &0.91& 16.78 & 16.00 & 14.59 & Oct. 06 2010  & 2400  &  A \\ 
GALEX J224029.0+120056.3 &22 40 29.01   &+12 00 55.7&19& 17.56 &0.85& 15.73 & 14.59 & 14.37 & Oct. 05 2010  & 1200  &  A \\
GALEX J224143.6+115326.2 &22 41 43.67   &+11 53 25.4&14& 17.08 &1.27& 14.27 & 14.18 & 13.91 & Oct. 14 2009  & 1200  &   A\\
 
  \noalign{\smallskip}\hline
\end{tabular}
\ec

\tablecomments{0.86\textwidth}{Column (2)-(3) list the right ascention (J2000.0) and corresponding declination (J2000.0) of the
objects given by 2MASS except GALEX J160902.8+524224.4 given by SDSS, column(4) lists the total number of detections
($NUV_{det}$) of the variable sources, column (5) lists the maximum NUV magnitude
for the source (measured in a single exposure) and column (6) lists the variation between the corresponding
maximum and minimum NUV magnitudes, column (7)-(9) list the photometric magnitudes of B2, R2 and I given by USNO-B1.0}
\end{table}

\section{RESULTS AND DISCUSSIONS}
\label{sect:results}

\subsection{Spectroscopic identification}

The observed 1-dimensional spectra are visually classied into various types. The spectral 
types are listed in Column (12) in Table 1.
All the observed sources are classified as stars except one AGN (i.e., GALEX\,J091332.6-101108.2). 
The spectrum of the object is shown in Figure 2. The spectrum of GALEX\,J091332.6-101108.2 shows a strong 
broad H$\alpha$ emission line, but a very weak and even undetectable broad H$\beta$ component, which 
allows us to classify the object as a Seyfert 1.9 galaxy rather than a typical type I AGN. 
The redshift is inferred to be z=0.055 according to the H$\alpha$ emission line. The 
continuum of the object is dominated by the stellar absorption features emitted from the host 
galaxy, which makes the object is not included in the quasar candidates selected 
from the GUVV2 catalog through their colors by Wheatley et al. (2008). 

   \begin{figure}[h!!!]
   \centering
   \includegraphics[width=9.0cm, angle=0]{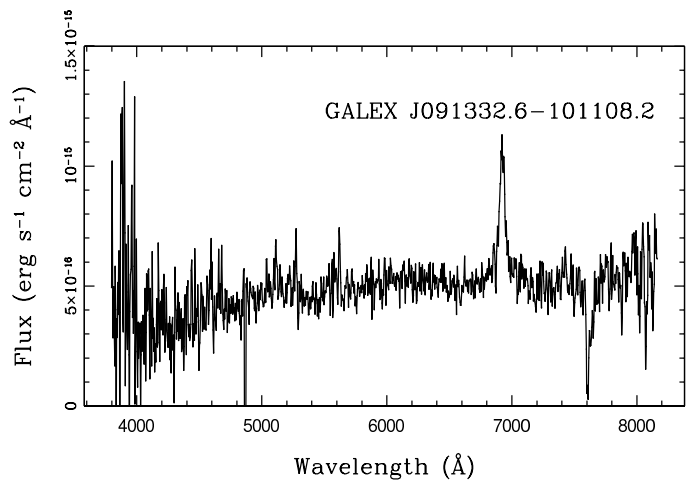}

   \begin{minipage}[]{85mm}

   \caption{ Spectrum of the newly identified AGN: GALEX\,J091332.6-101108.2  } \end{minipage}
   \label{Fig2}
   \end{figure}

GALEX NUV and FUV observation is found to be a sensitve probe for the variability 
of RR Lyare stars. The flux variability in RR Lyare stars is resulted from the radial pulsations 
that result in a cyclic temperature variation because of the stellar surface contraction and expansion. 
Out of the total 37 objects, 
19 objects show spectra typical of type -A to -F stars since their strong Balmer absorption features and 
Balmer jumps. A typical spectrum is presented in Figure 3 as an illustration. Figure 4 shows the
distribution of NUV magnitude variability for these identified type -A to -F stars.
All the type -A to -F stars have large NUV magnitude variability. The average and median vaules is 1.22 and 1.20 
magnitude, respectively, which is typical of the variability of RR Lyrae stars (Wheatley et al. 2005).

   \begin{figure}[h!!!]
   \centering
   \includegraphics[width=9.0cm, angle=0]{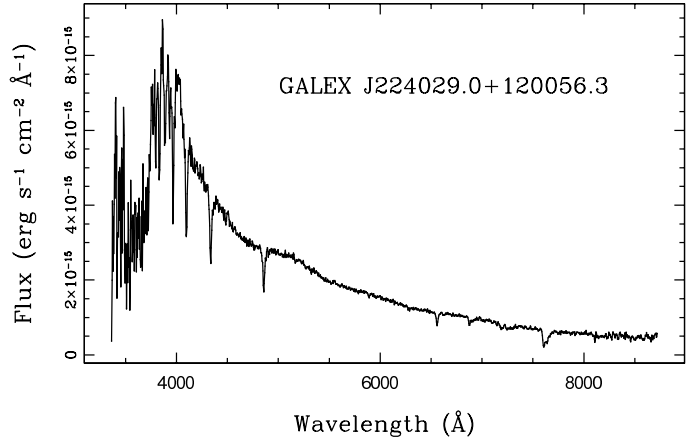}

   \begin{minipage}[]{85mm}

   \caption{ A typical spectrum of newly identified A-type star. } \end{minipage}
   \label{Fig3}
   \end{figure}

   \begin{figure}[h!!!]
   \centering
   \includegraphics[width=9.0cm, angle=0]{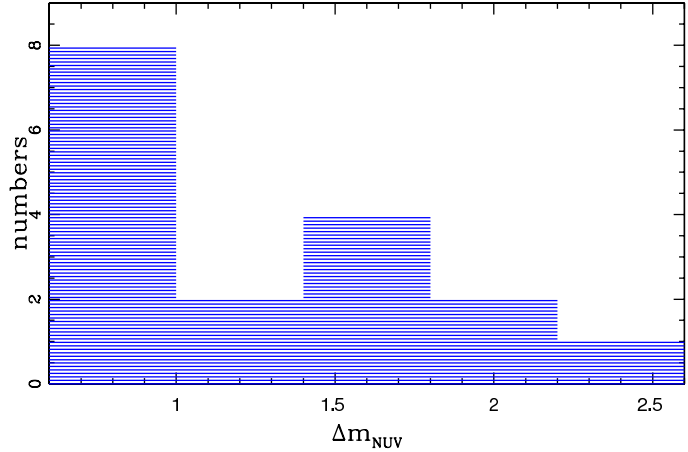}

   \begin{minipage}[]{85mm}

   \caption{ Distribution of the $\Delta m_{NUV}$ for the newly identified type -A to -F stars.} \end{minipage}
   \label{Fig4}
   \end{figure}

Seven M dwarf stars are identified in our spectroscopic observations. 
The spectra are present in Figure 5 for the 7 M dwarf stars.
Five of them are active 
during our observations because of the detected H$\alpha$ emission line.
M dwarf flare (dMe) stars are believed to be related with the coronal activity caused by their strong magnetic field 
coupled stellar disks (see review in Haisch et al. 1991). The flares last from seconds to hours in X-ray, 
ultraviolate, optical and radio
bands (e.g., Schmitt et al. 1993; Phillips et al. 1998; Stepanov et al. 1995; Welsh et al. 2006).
Although the physics of the flares is still poorly understood at present, it is widely believed that the UV flare 
emission in dMe is produced by the hot gas with temperature $10^5$K that are located at the chromosphere. 
In addtion to the flare events, the coronal activity of dMe stars could be alternatively spectroscopic 
identified according to their strong Balmer emission lines, especailly H$\alpha$ (e.g.,Worden \& Peterson 1976;
Shkolnik et al. 2011).

   \begin{figure}[h!!!]
   \centering
   \includegraphics[width=9.0cm, angle=0]{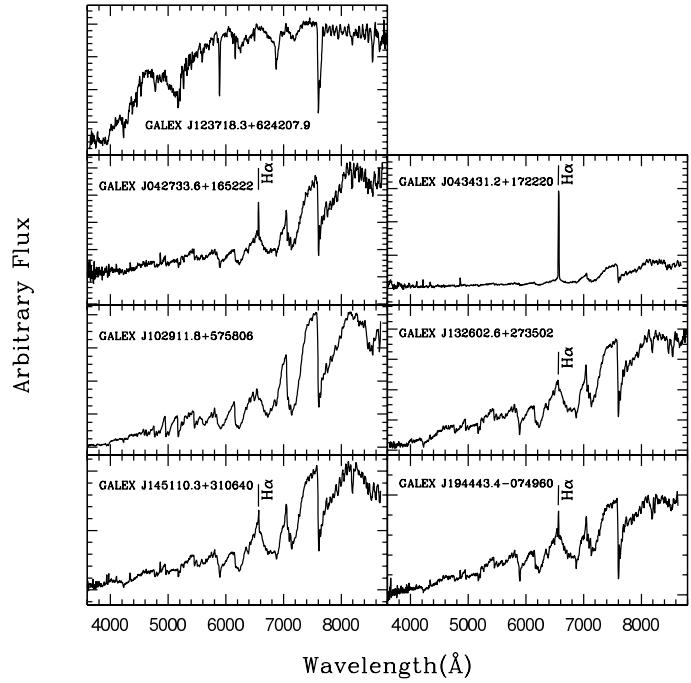}

   \begin{minipage}[]{85mm}

   \caption{ Spectra of seven newly discovered M dwarf stars} \end{minipage}
   \label{Fig5}
   \end{figure}

Different from the other six M dwarf stars, the source (GALEX\,J123718.3+624207.9) shows a spectra of extreme early
type ($\sim$M0) M dwarf star (see Fig.4).
We here attempt to quantatively determined the spectral types for the other six M dwarf stars by already comprehensively
studied spectral indices that are now widely used to produce quantative spectral classification for M dwarfs
arccording to their red part of the spectra (i.e., 6300-9000\AA, e.g., Reid et al. 1995; Kirkpatrick et al. 1995;
Martin et al. 1999; Hawley et al. 2002). Each index is defined as the ratio between the
mean flux in the pseudo-continuum bands and molecular band. Taking into account of the
spectral resolution of our spectra, the used spectral indices includes: TiO5, CaH2, CaH3 (Reid et al. 1995),
TiO7140 (Wilking et al. 2005), VO2 (L\'{e}pine et al. 2003), and C81(Stauffer et al. 1999).
The values of the spectral indices are calculated for all the objects by using the \it SPLOT \rm task in
IRAF package, and are listed in Table 2. Our spectral type determination indicates that the spectral
types calculated by different indices are generally consistent with each other within an uncertainty
of about 1 sub-type for individual object. The determined average spectral types are listed
in the last column in Table 2.

\begin{table}
\bc

\begin{minipage}[]{100mm}

\caption[]{M dwarf stars: spectral types and corresponding spectral indices}\end{minipage}

\small
 \begin{tabular}{lccccccccc}
  \hline\noalign{\smallskip}
GALEX Name & TiO 5  & CaH 2 & CaH 3 & TiO 7140  & VO 2 & C81 & Spectral type  \\
  \hline\noalign{\smallskip}
GALEX J042733.6+165222 & 0.415 & 0.411 & 0.681  & 1.783 & 0.746&1.275 &M3.78  \\
GALEX J043431.2+172220 & 0.456 & 0.441 & 0.724  & 1.807 & 0.685&1.400 &M3.70\\
GALEX J102911.8+575806 & 0.389 & 0.492 & 0.889  & 2.358 & 0.755&1.368 &M3.61\\
GALEX J132602.6+273502 & 0.535 & 0.510 & 0.787  & 1.555 & 0.832&1.160 &M2.63  \\
GALEX J145110.3+310640 & 0.531 & 0.446 & 0.713  & 1.646 & 0.816&1.286 &M3.22 \\
GALEX J194443.4-074960 & 0.499 & 0.490 & 0.735  & 1.495 & 0.892&1.107 &M2.71 \\
  \noalign{\smallskip}\hline
\end{tabular}
\ec

\end{table}

\subsection{Cyclic light curves}
Two RR Lyrae star candidates without spectroscopic identifications, GALEX\,J100133.2+014328.5 and 
GALEX\,J155800.6+535233.6, were monitored in optical V- and R-bands by us. Both light curves in R-band 
are presented in Figure 6. Cyclic variations can be identified from the light curves for both 
objects, although the light curves are under-sampling. 
For GALEX\,J100133.2+014328.5,
the light curve in the first night covers not only the peak, which is confirmed by 
the observation in the next day, but also the followed valley. The PDM method yields a period roughly at 0.54
day. Given the light curve obtained in the first night, the R-band flux variabilibty in the objects  
is as large as 1 mag, which is consitent with the typical value of RR Lyare stars (Skillen et al. 1993).
Although the valley in the light curve is covered twice in GALEX\,J155800.6+535233.6, 
the peak in the light curve is missed in our obervations because of the under-sampling. 
The poor sampling prevents us obtaining a period by the PDM method. 
The amplitude of NUV band of GALEX\,J100133.2+014328.5 and GALEX\,J155800.6+535233.6 are 1.98 and 0.87,
respectively, which are far larger than the amplitude of the optical band.
Combing the colors, cyclic light curves, and amplitude of ultroviolet and optical band
allows us to classify the two objects as RR Lyare stars which preliminarily proof the result of
the color-color diagram.

   \begin{figure}[h!!!]
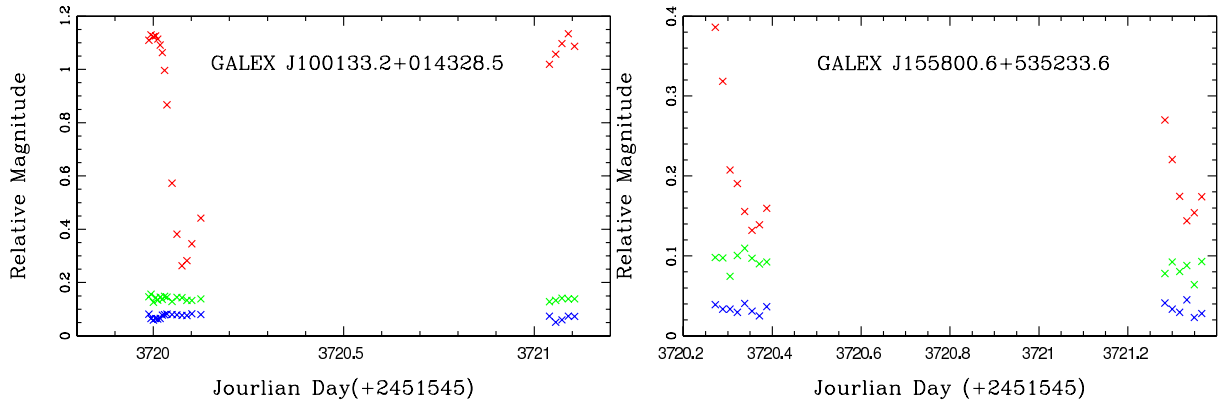

 \includegraphics[width=80mm]{J100133.epsi}
 \includegraphics[width=80mm]{J155800.epsi}
   \begin{minipage}[]{85mm}

   \caption{Light curves of two RR Lyrae star candidates.} \end{minipage}
   \label{Fig6}
   \end{figure}

\subsection{Color-color diagram}
In order to confirm the generality of the results derived from Figure 1,
We compare the newly spectroscopically identified objects with the previously identified ones in the 
NUV-optical color-color diagram (Fig. 7). 
The newly identified type -A to -F stars, type -G to -K stars and M dwarf stars
are marked as filled triangles with colors of green, yellow, and red, respectively.

The type -A to -F stars and M dwarf stars are clustered together with the previously identified RR Lyrae
stars and M dwarf stars, respectively, which show good consistence with former results from Figure 1.
Between the regions of type -A to -F stars and M dwarf stars, there gather some type -G to -K stars which 
may be solar-like coronal active stars, furtherly confirming the distribution trend of sources with
different energy spectrum.
The only one AGN marked as a large blue open star is classified as a Seyfert 1.9 galaxy, whose location
is consistent with its continuum dominated by the stellar absorption features emitted from the host galaxy.

   \begin{figure}[h!!!]
   \centering
   \includegraphics[width=9.0cm, angle=0]{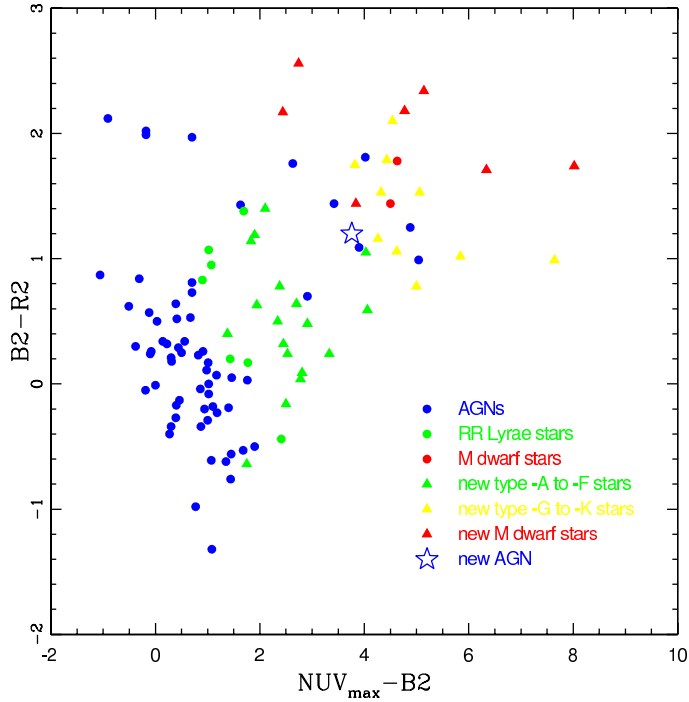}

   \begin{minipage}[]{85mm}

   \caption{ Plot of ($NUV_{max}- B2$) vs. (B2 -- R2) for the newly spectroscopically identified 
objects compared with the previously identified ones. The filled triangles with colors of green,
yellow and red are newly discovered type -A to -F stars, type -G to -K stars and M dwarf stars;
the large blue star is a newly discovered AGN. The previously identified sources are samely marked
as Fig.6.} 
   \end{minipage}
   \label{Fig7}
   \end{figure}

\subsection{Statistics of the identified sources}
There are totally 150 sources containing the newly spectroscopically identified and the sources
with previously identifications. The statistics of all the identified sources with different types are listed
in Table 3. 
\begin{table}
\bc

\begin{minipage}[]{100mm}

\caption[]{Statistics of all the identified sources}\end{minipage}

\small
 \begin{tabular}{ccccccc}
  \hline\noalign{\smallskip}
 AGNs&RR Lyrae stars/Type A-F stars&Type G-K stars& M dwarf stars&X-binaries& Others & Total  \\
 (1)  & (2)                           &(3)           & (4)          &   (5)    &  (6)   & (7)    \\  
  \hline\noalign{\smallskip}
    78  &    25                       & 10           &   9          &   19     &  9     & 150   \\
  52 \% &    16.7\%                   &    6.7\%     &    6\%       &   12.7\% & 6\%    & 100\% \\
  \noalign{\smallskip}\hline
\end{tabular}
\ec
\tablecomments{0.86\textwidth}{Column (2) lists the previously identified RR Lyraes together with the newly
identified type -A to -F stars.}

\end{table}

\section{Conclusions}
\label{sect:conclusion}

We have made a NUV-optical color-color plot for the previously known sources and the unidentified ones in GUVV2
, which separate sources into different regions. The identified sources in each region corresponds to a
certain spectral type, which allows us to tentatively classify the unknown sources simply by their
NUV and optical magnitudes.
In this method, we chosed two initially classified RR Lyrae star candidates to perform photometric monitors,
there NUV and optical amplitude together with the cyclic light curves of which allows us to finally classify
them as RR Lyrae stars.
We have also carried out spectroscopic observations for 37 randomly chosen unkown GUVV2 sources.
There are seven M dwarf flare stars and an AGN. Five of the M dwarf stars are active 
during our observations because of their H$\alpha$ emission. The AGN shows a strong 
broad H$\alpha$ emission line, but a very weak and even undetectable broad H$\beta$ component, which 
allows us to classify the object as a Seyfert 1.9 galaxy rather than a typical type I AGN.
The other sources are stars with spectral types from A to K. 
The location of all these newly identified sources on the NUV-optical diagram shows good consistence with
the previously known sources, which confirm the correctness and generality of the initially classifications.
Thus, at least for our results, we can rapidly and effectively classify the unknown ultroviolet variable
sources into several spectral types by their NUV and optical magnitudes in advance.

\normalem
\begin{acknowledgements}
JW and JYW are supported by National Basic Research Program of China (Grant 2009CB824800).
We would like to thank the anonymous referee for very useful comments and important
suggestions that improved the presentation.
This work was partically supported by the Open Project Program of the Key Laboratory
of Optical Astronomy, NAOC, CAS.
We thank Feng Qichen and Bai Yu for their share of observing time. 
Special thanks go to the staff at Xinglong
observatory as a part of National Astronomical Observatories, China Science Academy
for their instrumental and observational help.

\end{acknowledgements}

\label{lastpage}

\end{document}